\documentclass[conference]{IEEEtran}
\IEEEoverridecommandlockouts

\usepackage{graphicx}
\usepackage{subcaption}
\usepackage{epsfig}
\usepackage{mathptmx}
\usepackage{times}
\usepackage{amsmath}
\usepackage{amssymb}

\usepackage{cite}
\usepackage{color}

\graphicspath{{Figures/}}

\newtheorem{theorem}{Theorem}
\newtheorem{remark}{Remark}

\title{\LARGE \bf Distributed surveillance by a swarm of UAVs operating under positional uncertainty}

\author{\IEEEauthorblockN{Nikolaos Bousias}\IEEEauthorblockA{Dept. of Electrical and\\ Computer Engineering\\University of Patras\\Rio 26500, Greece\\up1020031@upnet.gr}\and\IEEEauthorblockN{Sotiris Papatheodorou}\IEEEauthorblockA{Dept. of Computing\\Imperial College London\\London SW7 2AZ, UK\\sotirisp@protonmail.com}\and\IEEEauthorblockN{Mariliza Tzes}\IEEEauthorblockA{Dept. of Electrical and Systems\\ Engineering (GRASP Lab)\\University of Pennsylvania\\Philadephia, PA 19104,U.S.A.\\mariliza.tze@gmail.com}\and\IEEEauthorblockN{Anthony Tzes}\IEEEauthorblockA{Dept. of Electrical and\\ Computer Engineering\\NYU AD\\Abu Dhabi 129188, U.A.E.\\anthony.tzes@nyu.edu}
}

\begin{document}
	
\maketitle
	
\thispagestyle{empty}
\pagestyle{empty}

\begin{abstract}
This article proposes a collaborative control framework for an autonomous aerial swarm tasked with the surveillance of a convex region of interest. Each Mobile Aerial Agent (MAA) is equipped with a Pan-Tilt-Zoom (PTZ) camera of conical FOV and suffers from sensor-induced positional uncertainty. By utilizing a Voronoi-free tessellation strategy and a gradient scheme, the heterogeneous swarm self-organizes in a distributed manner to monotonically achieve optimal collective visual coverage of the region of interest, both in terms of quality and total area. Simulation studies are offered to investigate the effectiveness of the suggested scheme.
\end{abstract}


\section{Introduction}
Mobile robot teams have several potential applications, one of the frequently
studied ones being area coverage problems.  Coverage problems can be broadly
categorized as static or sweep coverage. In static or blanket coverage
\cite{Abbasi_Aut2017,Pierson_IJRR2017} the objective of the mobile agents is a
static configuration at which some performance criterion is optimized. In
dynamic or sweep coverage problems \cite{Palacios_IEEETACR2016,Franco_EJC2015a}
the performance criterion is time--varying, resulting in the agents moving
constantly. Other ways to categorize coverage problems are based on the
properties of the region of interest
\cite{Kantaros_Automatica15,Alitappeh_SC2016}, of the dynamic model of the
mobile agents \cite{Sharifi_IEEETCST2015,Luna_CDC2010} or on the type of their
onboard sensors \cite{Stergiopoulos_ICRA14,Arslan_ICRA2016}. The most common
approach to coverage problems is geometric optimization \cite{Stergiopoulos_IEEETAC15} with other
proposed approaches being event--triggered control
\cite{Nowzari_Automatica2012}, game theory \cite{Ramaswamy_ACC2016}, annealing
algorithms \cite{Kwok_IEEETCST2011} and model predictive control
\cite{Mavrommati_IEEETR2018}.

An inherent characteristic of most positioning systems is the uncertainty in
their measurements. Some proposed solutions applied to mobile robots are
probabilistic methods \cite{Habibi_IEEETAC2016}, safe trajectory planning
\cite{Davis_IEEERAL2016} or the use of Voronoi--like partitioning schemes
\cite{Papatheodorou_IJARS2018,Tzes_LCSS2018}. In this article the positioning
uncertainty model is similar to the one used in
\cite{Papatheodorou_IJARS2018,Tzes_LCSS2018} but the approach followed differs.
Instead of employing a Voronoi--like space partitioning, the positioning
uncertainty is incorporated in the agents' sensing patterns and the sensed
space is partitioned using a Voronoi--free technique similar to
\cite{Papatheodorou_RAS2017}.

Aerial agents are a popular platform for area coverage tasks due to their high
mobility and versatility. The case of cameras with up to 3 translational and 3
rotational degrees of freedom has been examined in \cite{Schwager_IEEE2011} and
an algorithm for information exchange has also been developed. In this work
however the cameras are not allowed to zoom and the cameras' localization is
precise. Moreover, regions covered by multiple agents contribute more to the
objective, thus favoring overlapping between the agents' sensed regions.
Previous works have examined downwards facing cameras
\cite{Papatheodorou_RAS2017,Papatheodorou_CDC2017} and although positioning
uncertainty has been successfully incorporated in these control schemes
\cite{Tzes_LCSS2018}, it was done using a Voronoi--like partitioning which is
not easy to generalize in the case of pan-tilt-zoom cameras. Pan-tilt-zoom
camera networks have been examined using Voronoi--like diagrams in
\cite{Arslan_ICRA2018} although in that work the cameras were stationary
instead of being affixed on mobile agents. In the present work the MAAs have 3
translational degrees of freedom and are equipped with pan-tilt-zoom cameras.
The MAAs' planar positioning uncertainty is taken into account by using a
Voronoi--free partitioning scheme. Additionally, regions sensed by multiple
agents do not contribute more to the objective, thus favoring separation of the
MAAs' sensing patterns. It should be noted that what the best approach
concerning the overlapping of sensing patterns depends entirely on the intended
use--case and the one used in this article can not be considered strictly
better or worse than the one used in \cite{Schwager_IEEE2011}.

\section{Problem Statement}
We assume a compact convex region $\Omega \in \mathbb{R}^2$ to be placed under
surveillance by a swarm of $n$ MAAs, each positioned at
$X_i=\left[x_i,y_i,z_i\right]^T, i\in I_n$ where $I_n=\{1, \dots, n\}$. We define
the vector $q_i = \left[x_i,y_i\right]^T \in \Omega$ denoting the projection of
each MAA on the plane. Each MAA can fly within a predefined altitude range,
thus $z_i \in \left[ z_i^{\min},z_i^{\max} \right], i\in I_n$ with $z_i^{\min}
\leq z_i^{\max}$. These altitude constraints ensure the safe operation of the
MAA by avoiding collisions with ground obstacles as well as ensuring they
remain within communication range of their base stations.

In addition, each MAA comes equipped with an onboard visual sensor capable of
pan and tilt movements. Moreover, the sensor has a conical field of view and is
able to alter its zoom. We denote the sensor's pan and tilt angles $h_i$ and
$\theta_i$ respectively, while its zoom level is represented by the angle of of
the cone of vision which is denoted $2 \delta_i$ with $\delta_i \in
\left[\delta_i^{\min} , \delta_i^{\max} \right]$ and $\delta_i^{\min} \leq
\delta_i^{\max} < \frac{\pi}{2}$.

Given the conical field of view of the sensors, its intersection with the plane
will be a conic section which we call the sensing pattern. The sensing pattern
is the region of the plane an MAA is able to cover. More specifically, it is a
circle for $h_i = 0$, an ellipse for $0 < \left|h_i\right| < \frac{\pi}{2} -
\delta_i$, a parabola for $\frac{\pi}{2} - \delta_i \leq \left|h_i\right| \leq
\frac{\pi}{2} + \delta_i$ and a hyperbola for $\frac{\pi}{2} + \delta_i <
\left|h_i\right|$. In the sequel we will examine only the case where
$\left|h_i\right| < \frac{\pi}{2} - \delta_i$, i.e.  circular and elliptical
sensing patterns, in order to always have the sensing pattern bounded by a
curve. In order to have static boundaries for the tilt angle we will constrain
it inside the interval $\left(-h_i^{\max}, h_i^{\max} \right) \subseteq
\left(-\frac{\pi}{2} + \delta_i, \frac{\pi}{2} - \delta_i \right)$ where
$h_i^{\max} = \frac{\pi}{2} - \delta_i^{\max}$.

We define the center of the sensing footprint $q_{c,i} \in \Omega$ and denote
the semi-major and semi minor axis of the elliptic sensing pattern $a_i$ and
$b_i$ respectively. The unit vector $w_i \in \mathbb{R}^2$ indicates the
orientation of the sensing pattern and is parallel to the semi-major axis of
the ellipse if the sensing pattern is elliptical. These result in each MAA's
sensing pattern being
\begin{equation}
	\label{eq:sensing}
	{C_i^s} (X_i,h_i,\theta_i,\delta_i) = \mathbf{R}\left(\theta_i\right) C_i^b
	+ q_{i,c}, \quad i \in I_n,
\end{equation}
where $\mathbf{R}$ is the ${2\times2}$ rotation matrix, $||.||$ is the
Euclidean metric and
\begin{align}
	C_i^b &= \left\{ q \in \mathbb{R}^2 \colon \left\Vert \left[ \begin{matrix}
	\frac{1}{a_i} & 0 \\ 0 & \frac{1}{b_i} \end{matrix} \right] {q} \right\Vert
	\leq 1 \right\}, \\
	a_i &= \frac{z_i}{2} \left[ \tan(h_i + \delta_i) - \tan(h_i - \delta_i)
	\right], \\  
	b_i &= {z_i} \tan(\delta_i) \sqrt{1 + \left[ \frac{ \tan(h_i + \delta_i) +
	\tan(h_i - \delta_i)}{2} \right]^2}, \\  
	{q_{i,c}} &= {q_i} + w_i \frac{z_i}{2} \left[ \tan(h_i + \delta_i) +
	\tan(h_i - \delta_i) \right], \\  
	w_i &= \begin{bmatrix} \cos(\theta_i) & \sin(\theta_i) \end{bmatrix}^T.
\end{align}
The pan angle $\theta_i$ only affects the sensing pattern's orientation, while
the tilt angle $h_i$ affects the eccentricity of the elliptical sensing
pattern. It can be shown that if $h_i = 0$ then $a_i = b_i$ and $C_i^s$
degenerates into a circle with $q_{c,i} = q_i$.

For the sake of simplicity, instead of using a complete dynamic model for the
MAAs such as quadrotor dynamics, a single integrator kineamtic model is used
instead. The MAAs are approximated by point masses able to move in
$\mathbb{R}^3$. It is assumed that the visual sensors' pan and tilt angles and
zoom can be controlled by onboard servos, thus their states are decoupled from
those of the MAA. Therefore the kinematic model of each MAA is
\begin{align}
	\label{eq:kin_q}
	\dot{q_i} &= u_{i,q}, \quad q_i \in \Omega, & u_{i,q} \in \mathbb{R}^2, \\  
	\label{eq:kin_z}
	\dot{z_i} &= u_{i,z}, \quad z_i \in \left[ z_i^{\min}, z_i^{\max} \right],
			  & u_{i,z} \in \mathbb{R}, \\  
	\label{eq:kin_theta}
	\dot{\theta_i} &= u_{i,\theta}, \quad \theta_i \in \mathbb{R}, & u_{i,\theta}
	\in \mathbb{R}, \\  
	\label{eq:kin_h}
	\dot{h_i} &= u_{i,h}, \quad h_i \in \left( -h_i^{\max}, h_i^{\max} \right),
			  & u_{i,h} \in \mathbb{R}, \\  
	\label{eq:kin_delta}
	\dot{\delta_i} &= u_{i,\delta}, \quad \delta_i \in \left[\delta_i^{\min},
	\delta_i^{\max} \right], & u_{i,\delta} \in \mathbb{R}.
\end{align}

The projection on the ground $q_i \in \Omega$ of each agent’s position is
assumed to be known with a degree of uncertainty, whereas each MAA's altitude
$z_i$, sensor pan angle $\theta_i$, tilt angle $h_i$ and view angle $\delta_i$
are known with certainty. Given an upper bound $r_i$ for the positioning
uncertainty of each MAA, its footprint $q_i$ may reside anywhere with a disk
called the positioning uncertainty region. The positioning uncertainty region,
denoted $C_i^u$, is defined as
\begin{equation}
	\label{eq:uncertainty}
	{C_i^u} (q_i, r_i) = \Big\{  q \in \Omega : \left\Vert q-q_i  \right\Vert
	\leq r_i \Big\}, \quad i \in I_n.
\end{equation}

\begin{figure}[htbp]
	\centering
	\includegraphics[trim=73 0 65 110,clip,width=0.42\textwidth]{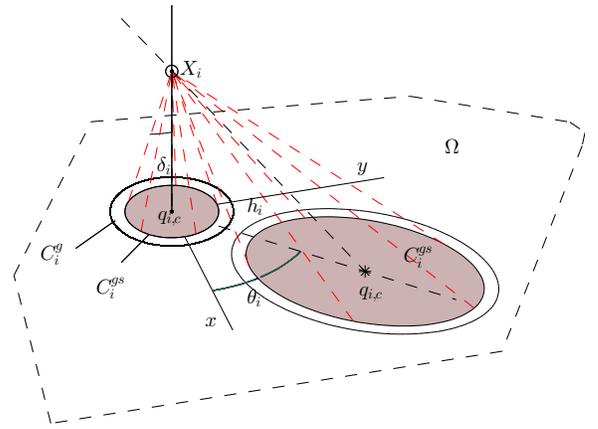}
	\caption{Visual coverage concept.}
	\label{fig:problem_statement}
\end{figure}

Given the positioning uncertainty of each MAA, we also define the
guaranteed sensed region $C_i^{gs} \subseteq C_i^s \subset \mathbb{R}^2$ as the
region the MAA is guaranteed to cover for all its possible positions within
$C_i^u$. The guaranteed sensed region is then defined as
\begin{align}
	\nonumber
	{C_i^{gs}} (X_i,h_i,\theta_i,\delta_i,r_i) &\stackrel{\triangle}{=} \left\{
	\underset{q_i \in C_i^u}{\bigcap} {C_i^s} \right\} \\ 
	\label{eq:gsensing}
	&= \mathbf{R}\left(\theta_i\right)C_i^{b_{gs}} + q_{i,c}, \quad i \in I_n,
\end{align} 
where
\begin{align}
	C_i^{b_{gs}}= \left\{ q \in \Omega \colon \left\Vert \left[  \begin{matrix}
	\frac{1}{a_i-r_i} & 0 \\ 0 &  \frac{1}{b_i-r_i} \end{matrix} \right] {q}
	\right\Vert \leq 1 \right\}.
\end{align}
Since the positioning uncertainty region $C_i^u$ and sensed region $C_i^s$ are
circular and elliptical respectively, the guaranteed sensed region $C_i^{gs}$
is also an ellipse. If $C_i^s$ is a disk due to the tilt angle $h_i$ being $0$
then $C_i^{gs}$ will also be a disk. If $r_i=0$, i.e. the position of the MAA's
footprint is known precisely, then $C_i^{gs} = C_i^s$. As $r_i \to
\min(a_i,b_i)$ then $ C_i^gs$ approaches a line segment. For $r_i >
\min(a_i,b_i)$ we get that $C_i^{gs}=\emptyset$. Figure
\ref{fig:problem_statement} illustrates the visual coverage concept described
in this section. The agent's guaranteed sensing pattern is shown filled in red
for both $h_i = 0$ and $h_i \in \left( 0, \frac{\pi}{2} - \delta_i \right)$.

Due to the nature of visual sensors, objects further away from the sensor
appear with lower quality than ones near the sensor. We model the coverage
quality using a quality function $f_i \colon \left[z_i^{\min}, +\infty\right]
\rightarrow \left[0, 1\right]$, with $0$ and $1$ corresponding to the lowest
and the highest possible quality respectively. In order to keep the control
scheme simple, it is assumed that the coverage quality is uniform throughout
each MAA's sensed pattern. As the MAA's altitude $z_i$ increases, the visual
coverage quality of its sensed region decreases.  The same is true while the
sensor's tilt angle $h_i$ increases, resulting in the center $q_{i,c}$ of the
sensed pattern $C_i^s$ moving further away from the MAA. Similarly, zooming
out, i.e. increasing $\delta_i$, also leads in a decrease in quality. However,
the sensor's pan angle $\theta_i$ as well as the agent's footprint $q_i$ have
no effect on the coverage quality. Except from being a decreasing function of
$z_i$, $h_i$ and $\delta_i$, $f_i$ must also be first--order differentiable
with respect to $z_i$, $h_i$ and $\delta_i$ within $C_i^{gs}$. This property is
needed for computing the control law and will become apparent in the sequel.

Although any function having the previously mentioned properties could be
chosen as the coverage quality function $f_i$, the following one was chosen
arbitrarily
\begin{small}
\begin{equation}
	\label{eq:quality}
	f_i (z_i,h_i,\delta_i) =
	\begin{cases} 
		\frac{p\left(z_i, \ z_i^{\min}, \ z_i^{\max}\right)}{3} +
		\frac{p\left(h_i, \ 0, \ h_i^{\max}\right)}{3} &\\
		\quad + \frac{p\left(\delta_i, \ \delta_i^{\min}, \
		\delta_i^{\max}\right)}{3}, &\ q \in C_i^{gs} \\
		0, &\ q\notin C_i^{gs} 
	\end{cases},
\end{equation}
\end{small}
where
\begin{equation}
	p\left(x, \ x^{\min}, \ x^{\max}\right) = \frac{\left(\left(x -
			x^{\min}\right)^2 - \left(x^{\max} -
	x^{\min}\right)^2\right)^2}{\left(x^{\max} - x^{\min}\right)^4}.
\end{equation}
For the function $p$ it holds that $p\left(x^{\min}, \ x^{\min}, \
x^{\max}\right) = 1$ and that $p\left(x^{\max}, \ x^{\min}, \ x^{\max}\right) =
0$. Consequently $f_i (z_i^{\min}, 0, \delta_i^{\min}) = 1$ and $\underset{h_i
\to h_i^{\max}}{\lim} f_i (z_i^{\max}, h_i, \delta_i^{\max}) = 0$. The function
$f_i$ is also dependent on the agent's altitude, tilt and zoom constraints. It
should be noted that this choice of quality function is not unique and that
different quality functions result in different quality--coverage trade--offs.

Additionally, each point $q\in \Omega$ can be assigned an importance weight
through a space density function $\phi \colon \Omega \rightarrow \mathbb{R}_+$
which expresses the a priori information regarding the importance of certain
regions of $\Omega$. We define the following joint coverage-quality objective
\begin{equation}
	\label{eq:objective_initial}
	\mathcal{H} \stackrel{\triangle}{=}  \int_{\Omega}{ \ \max_{\substack{i \in
	I_n}} f_i (z_i,h_i,\delta_i) \ \phi\left(q\right) \mathrm{d}q}.
\end{equation}
This function accounts for both the area covered by the agents and the coverage
quality over that area, while also taking into account the importance of points
as encoded by $\phi(q)$. The goal of the MAA team is to maximize this
objective. To that extent, a suitable partitioning scheme will be employed in
order to distribute the computation of $\mathcal{H}$ among the agents. Then a
gradient--based control law will be designed to lead the MAA team to a locally
optimal configuration with respect to $\mathcal{H}$.

\section{Area Partitioning Strategy}
The most common choice of partitioning scheme for area coverage problems is the
Voronoi diagram and similar diagrams inspired by it. Voronoi--like diagrams
that can take into account the positioning uncertainty of mobile agents have
been proposed in the past \cite{Papatheodorou_IJARS2018,Tzes_LCSS2018}. However
in this work a partitioning of just the sensed space is utilized, similarly to
\cite{Papatheodorou_CDC2017}. This partitioning scheme assigns a region of
responsibility (cell) to each agent based on guaranteed sensed regions
$C_i^{gs}$ and the coverage quality over them. Each MAA is assigned a cell
$W_i$ as follows
\begin{equation}
	\label{eq:partitioning}
	W_i \stackrel{\triangle}{=} \left\{ q \in \Omega \colon \ f_i > f_j, \ i
	\neq j \right\}, i \in I_n.
\end{equation}
However the union these cells does not comprise a complete tessellation of the
total guaranteed sensed region $\underset{i \in I_n}{\bigcup} C_i^{gs}$. This
is due to the fact that regions sensed by multiple agents with the same
coverage quality are left unassigned. These so called common regions still
contribute towards the objective $\mathcal{H}$ so they must be taken into
account. The set of agents with the same coverage quality $f^l$ and overlapping
guaranteed sensed regions is
\begin{align*}
	\mathcal{I}_l = &\left\{ i,j \in I_n, i \neq j \colon C_i^{gs} \cap
		C_j^{gs} \neq \emptyset
	\right.\\
	&\quad\left.\wedge  f_i  = f_j  = f^l \right\}, l \in I_L.
\end{align*}
The common regions are then computed as
\begin{align}
	\label{eq:partitioning_common}
	W_c^l = \left\{ \exists i,j \in \mathcal{I}_l, i \neq j:q\in C_i^{gs} \cap
	C_j^{gs} \right\},\quad l \in I_L.
\end{align}

By utilizing the partitioning strategy \eqref{eq:partitioning},
\eqref{eq:partitioning_common}, coverage--quality objective
\eqref{eq:objective_initial} can be written as
\begin{equation}
	\label{eq:objective}
	\mathcal{H} = \sum_{\substack{i\in I_n}}
	\int_{\substack{W_i}}{f_i \ \phi(q) \
	\mathrm{d}q} + \sum_{l=1}^{L} \int_{\substack{W_c^l}}{ f^l
	\ \phi(q) \ \mathrm{d}q} 
\end{equation}

\begin{remark}
	We define the neighbors $N_i$ of an agent $i$ as
	\begin{equation}
		\label{eq:neighbors}
		N_i \stackrel{\triangle}{=} \left\{j \in I_n \setminus i \colon
		C_i^{gs} \cap C_j^{gs} \neq \emptyset \right\}.
	\end{equation}
	The neighbors of an agent $i$ are essentially the agents that affect the
	cell $W_i$ of agent $i$, thus they are the agents $i$ must be able to
	exchange information with.

	By allowing the MAAs' cameras to tilt it is possible that the sensed
	regions of two distant MAAs overlap. Since the partitioning scheme is based
	on the sensed regions, these MAAs should be able to communicate. However
	this might not always be practical given their distance. An algorithm for
	propagating state information in a mobile agent network has been proposed
	in \cite{Schwager_IEEE2011}. By utilizing this algorithm MAAs are able to
	exchange information with their neighbors in multiple hops instead of
	communicating directly.
\end{remark}

\begin{remark}
	Since the partitioning scheme \eqref{eq:partitioning},
	\eqref{eq:partitioning_common} only partitions the guaranteed sensed region
	$\underset{i \in I_n}{\bigcup} C_i^{gs}$, a portion of $\Omega$ is left
	unpartitioned. This region is called the neutral region, is denoted
	$\mathcal{O}$ and can be computed as
	\begin{align}
		\label{eq:neutral}
		\mathcal{O} &= \Omega \setminus \left\{ \underset{i \in I_n}{\bigcup}
		{W_i} \cup \underset{i \in I_l}\bigcup{W_c^l} \right\}= \Omega \setminus \underset{i \in I_n}{\bigcup} C_i^{gs}.
	\end{align}
\end{remark}

\begin{remark}
	Due to the fact that the coverage quality $f_i$ is constant throughout the
	guaranteed sensed region $C_i^{gs}$, the resulting cells $W_i$ are bounded
	by elliptical arcs of $C_i^{gs}$. Moreover, this partitioning scheme may
	result in some cells being non--convex, empty or consisting of multiple
	disjoint regions. However all of these cases are handled properly by the
	designed control law without the need for extensions.
\end{remark}

\section{Collaborative Control Development}
Having defined the partitioning scheme \eqref{eq:partitioning},
\eqref{eq:partitioning_common} which allows distributing the computation of the
objective $\mathcal{H}$ among the agents, what remains is the derivation of the
gradient--based control law.
\begin{theorem}
	Given a team of MAAs with kinematics described by \eqref{eq:kin_q},
	\eqref{eq:kin_z}, \eqref{eq:kin_theta}, \eqref{eq:kin_h},
	\eqref{eq:kin_delta}, sensing performance \eqref{eq:sensing} and
	positioning uncertainty \eqref{eq:uncertainty}, the following control law
	guarantees monotonic increase of the coverage--quality objective
	\eqref{eq:objective} along the MAAs trajectories.
	\begin{small}
	\begin{eqnarray}
		\label{eq:control_q}
		u_{i,q} &= K_{i,q} \Biggl[ \underset{\partial W_i \cap \partial
		\mathcal{O}} {\int} {u_i^i \ n_i \ f_i \ \phi(q) \ \mathrm{d}q} \qquad\qquad\qquad\qquad\qquad \nonumber \\  
		&\quad\quad +\sum_{\substack{j \in I_n \\ j \neq i}} \underset{\partial W_i \cap
		\partial W_j}{\int}{u_i^i \ n_i \left( f_i -f_j \right) \ \phi(q) \
		\mathrm{d}q} \Biggr], \\ 
		\label{eq:control_z}
		u_{i,z} &= { K_{i,z} \Biggl[ \underset{\partial W_i \cap \partial
		\mathcal{O}}{\int} {v_i^i \ n_i \ f_i \ \phi(q) \ \mathrm{d}q} + {
		\underset{W_i}{\int}{ \frac{\partial f_i} {\partial z_i}  \phi(q) \ \mathrm{d}q}} } \quad\quad \nonumber \\  
		&\quad\quad +\sum_{\substack{j \in I_n \\ j \neq i}} \underset{\partial W_i \cap
		\partial W_j}{\int}{v_i^i \ n_i \left(f_i- f_j \right) \ \phi(q) \
		\mathrm{d}q} \Biggr], \\  
		\label{eq:control_theta}
		u_{i,\theta} &= K_{i,\theta} \Biggl[ \underset{\partial W_i \cap
		\partial \mathcal{O}}{\int} {\tau_i^i \ n_i \ f_i \ \phi(q) \ \mathrm{d}q} \qquad\qquad\qquad\qquad\qquad \nonumber \\  
		&\quad\quad + \sum_{\substack{j \in I_n \\ j \neq i}} \underset{\partial W_i \cap
		\partial W_j}{\int}{\tau_i^i \ n_i \left( f_i- f_j \right) \ \phi(q) \
		\mathrm{d}q} \Biggr], \\  
		\label{eq:control_h}
		u_{i,h} &={ K_{i,h} \Biggl[ \underset{\partial W_i \cap \partial
		\mathcal{O}}{\int} {\sigma_i^i \ n_i \ f_i \ \phi(q) \ \mathrm{d}q} +
		\underset{W_i}{\int}{ \frac{\partial f_i} {\partial h_i} \ \phi(q) \ \mathrm{d}q} } \quad\quad \nonumber \\  
		&\quad\quad {+ \sum_{\substack{j \in I_n \\ j \neq i}} \underset{\partial W_i
		\cap \partial W_j}{\int}{\sigma_i^i \ n_i \left(  f_i- f_j \right) \
		\phi(q) \ \mathrm{d}q} } \Biggr], \\  
		\label{eq:control_delta}
		u_{i,\delta} &={ K_{i,\delta} \Biggl[ \underset{\partial W_i \cap
		\partial \mathcal{O}}{\int} {\mu_i^i \ n_i \ f_i \ \phi(q) \ \mathrm{d}q}
		+\underset{W_i}{\int}{ \frac{\partial f_i} {\partial \delta_i}  \ \phi(q)   \: \mathrm{d}q}  } \quad\quad \nonumber \\  
		&\quad\quad {+ \sum_{\substack{j \in I_n \\ j \neq i}} \underset{\partial W_i
		\cap \partial W_j}{\int}{\mu_i^i \ n_i \left( f_i- f_j \right) \
		\phi(q) \ \mathrm{d}q} } \Biggr],
	\end{eqnarray}
	\end{small}
	where $K_{i,q},K_{i,z},K_{i,\theta},K_{i,h},K_{i,\delta}$ are positive
	constants, $n_i$ the outward pointing unit normal vector on $W_i$ and
	$u_i^i,v_i^i,\tau_i^i,\sigma_i^i,\mu_i^i$ the Jacobian matrices
	\begin{eqnarray}
		u_j^i &\stackrel{\triangle}{=}  \dfrac{\partial q}{\partial q_i},\quad q \in \partial W_j,i,j\in I_n \\  
		v_j^i &\stackrel{\triangle}{=}  \dfrac{\partial q}{\partial z_i},\quad q \in \partial W_j,i,j\in I_n \\  
		\tau_j^i &\stackrel{\triangle}{=}  \dfrac{\partial q}{\partial \theta_i},\quad q \in \partial W_j,i,j\in I_n \\  
		\sigma_j^i &\stackrel{\triangle}{=}  \dfrac{\partial q}{\partial h_i},\quad q \in \partial W_j,i,j\in I_n \\  
		\mu_j^i &\stackrel{\triangle}{=}  \dfrac{\partial q}{\partial \delta_i},\quad q \in \partial W_j,i,j\in I_n
	\end{eqnarray}
\end{theorem}

\begin{IEEEproof}
In order to guarantee monotonic increase of $\mathcal{H}$, its time derivative
is evaluated as
\begin{equation}
	\dfrac{\partial \mathcal{H}} {\partial t} =  \dfrac{\partial \mathcal{H}} {\partial q_i} \dot{q_i} + \dfrac{\partial \mathcal{H}} {\partial z_i} \dot{z_i}+ \dfrac{\partial \mathcal{H}} {\partial \theta_i} \dot{\theta_i}+ \dfrac{\partial \mathcal{H}} {\partial h_i} \dot{h_i}+ \dfrac{\partial \mathcal{H}} {\partial \delta_i} \dot{\delta_i}
\end{equation}
By selecting the following control inputs
\begin{align*}
	u_{i,q} &= K_{i,q} \dfrac{\partial \mathcal{H}} {\partial q_i}, u_{i,z} = K_{i,z} \dfrac{\partial \mathcal{H}} {\partial z_i} ,  \nonumber\\ 
	u_{i,\theta} &= K_{i,\theta} \dfrac{\partial \mathcal{H}} {\partial \theta_i} , u_{i,h} = K_{i,h} \dfrac{\partial \mathcal{H}} {\partial h_i} ,u_{i,\delta} = K_{i,\delta} \dfrac{\partial \mathcal{H}} {\partial\delta_i},	
\end{align*}

we guarantee, given the MAAs’ dynamics, that $\dfrac{\partial \mathcal{H}}
{\partial t} $ is non-negative since
\begin{eqnarray}
	&\dfrac{\partial \mathcal{H}} {\partial t} = \Bigg[  K_{i,q} \left( \dfrac{\partial \mathcal{H}} {\partial q_i} \right)^2 +K_{i,z} \left( \dfrac{\partial \mathcal{H}} {\partial z_i} \right)^2 + K_{i,\theta} \left( \dfrac{\partial \mathcal{H}} {\partial \theta_i} \right)^2  \qquad\qquad \qquad\qquad  \nonumber\\  
	&+ K_{i,h} \left( \dfrac{\partial \mathcal{H}} {\partial h_i}+ \right)^2
+K_{i,\delta} \left( \dfrac{\partial \mathcal{H}} {\partial \delta_i} \right)^2
\Bigg] \geq 0, \nonumber
\end{eqnarray}
where $K_{i,q},K_{i,z},K_{i,\theta},K_{i,h},K_{i,\delta}$ are positive
constants, thus, ensuring that the coverage--quality criterion increases in a
monotonic manner.

The partial derivative $\frac{\partial \mathcal{H}}{\partial q_i}$ is
\begin{small}
\begin{equation*}
	\dfrac{\partial \mathcal{H}}{\partial q_i} =  \dfrac{\partial}{\partial
	q_i} \left\{ \sum_{\substack{i\in I_n}}  \underset{W_i}{\int}{ f_i \
\phi(q) \ \mathrm{d}q} + \sum_{l=1}^{L} \underset{W_c^l}{\int}{ f^l \ \phi(q) \
\mathrm{d}q}   \right\}.
\end{equation*}
\end{small}
By applying the Leibniz integral rule and since $\frac{\partial f_i
(z_i,h_i,\delta_i)}{\partial q_i}=\frac{\partial f_j(z_j,h_j,\delta_j)
}{\partial q_i}=0$ the previous equation yields
\begin{align*}
	\dfrac{\partial \mathcal{H}} {\partial q_i} &=  \underset{\partial
	W_i}{\int} {u_i^i \ n_i \ f_i \ \phi(q) \ \mathrm{d}q} \\  
	&\quad + \sum_{\substack{j \in I_n \\ j \neq i}} \underset{\partial W_i \cap
	\partial W_j}{\int}{u_j^i \ n_j \ f_j \phi(q) \ \mathrm{d}q}.
\end{align*}
We use a boundary decomposition of $\partial W_i$ into disjoint sets similarly
to \cite{Papatheodorou_CDC2017}
\begin{align}
	\partial W_i &= \Bigg\{  \left\{ \partial W_i \cap \partial \Omega \right\} \cup \left\{ \partial W_i \cap \partial \mathcal{O} \right\} \cap \qquad\qquad\qquad \nonumber \\  
	&\qquad \Bigg\{ \bigcup_{\substack{i \neq j}}{\partial W_i \cap \partial W_j}  \Bigg\} \cap \Bigg\{ \bigcup_{l=1}^{L}{\partial W_i \cap \partial W_c^l}  \Bigg\}  \Bigg\}
\end{align}
and assuming a static region of interest $\Omega$. In addition, since $\partial
W_i\cup \partial W_c^l$ are subsets of some sensed region boundary $\partial
C_j$, independent of the state of node $i$, at $q \in \bigg\{ \partial \Omega
	\cap \partial W_i \bigg\}$ and $q \in \bigg\{ \partial W_i \cup \partial
W_c^l \bigg\}$, the Jacobian matrix is $u_i^i=\bf0_{2\times2}$ resulting in the
final expression for $\dfrac{\partial \mathcal{H}} {\partial q_i}$
\begin{align*}
	\dfrac{\partial \mathcal{H}} {\partial q_i} &= \underset{\partial W_i \cap
\partial \mathcal{O}}{\int} {u_i^i \ n_i f_i \ \phi(q) \ \mathrm{d}q}  \\  
	&\qquad + \sum_{\substack{j \in I_n \\ j \neq i}} \underset{\partial W_i \cap
	\partial W_j}{\int}{u_i^i \ n_i \left[f_i - f_j \right] \ \phi(q) \
\mathrm{d}q}.
\end{align*}

Through a similar procedure and given that $\frac{\partial f_j}{\partial
z_i}=\frac{\partial f_j}{\partial \theta_i}=\frac{\partial f_i}{\partial
\theta_i}=\frac{\partial f_j}{\partial h_i}=\frac{\partial f_j}{\partial
\delta_i}=0$ we obtain the rest of the control laws.
\end{IEEEproof}

\section{Simulation Studies}
Simulation studies were conducted in order to evaluate the efficiency of the
proposed control strategy. For consistency, the region of interest $\Omega$ was
selected to be the same as in \cite{Tzes_LCSS2018}. The space density function
was assumed to be $\phi \left(q\right)=1,\forall q \in \Omega$, assigning equal
importance to all points inside the region of interest. The camera state limits
were $h_i^{\max} = 50^\circ$, $\delta_i^{\min} = 15^\circ$ and $\delta_i^{\max}
= 35^\circ, \ \forall i \in I_n$. The MAAs’ cells are shown filled in grey with
solid black boundaries while the boundaries of guaranteed sensed regions are
shown in dashed red lines in Figures 2 (a) and (b) and Figures 4 (a) and (b).
The MAAs' trajectories are shown as blue lines Figures 2 (b) and 4 (b).

\subsection{Case study I}
This simulation examines the case of a team of 3 MAAs and it serves to
highlight the fact that there exists a configuration with respect to the
agents' altitude $z_i$, camera tilt angle $h_i$ and zoom $\delta_i$ that is
globally optimal. The MAAs altitude constraints were set to $ z_i^{min}=0.3$
and $z_i^{max}=3.8 \ \forall i \in I_3$. The initial and final configurations
of the swarm are shown in Figures 2 (a) and 2 (b)
respectively. We observe from Figure 2 (b) that the MAAs
guaranteed sensed regions do not overlap and are completely contained inside
$\Omega$. This indicates that the MAA team has reached a globally optimal
configuration.  It should be noted that due to the fact that no closed--form
expression exists for the arc length of an ellipse, it is impossible to
analytically compute the altitude, tilt angle and zoom that result in this
configuration. This simulation study was also conducted with agents equipped
with downward facing cameras unable to pan, tilt or zoom as in
\cite{Papatheodorou_RAS2017}. Figure 3 shows the evolution of
the coverage--quality objective $\mathcal{H}$ over time for both the
pan-tilt-zoom and downwards facing cameras in solid black and dashed red
respectively. It is observed that by allowing the MAAs' cameras to pan, tilt
and zoom, it is possible to achieve significantly higher coverage performance.
Finally, it is observed that the monotonic increase of $\mathcal{H}$ has been
achieved, confirming that the control design and implementation is correct.

\begin{figure}[htbp]
	\centering
	\begin{subfigure}[b]{0.5\textwidth}
		\centering
		\includegraphics[trim=10 25 10 20,clip,scale=0.5]{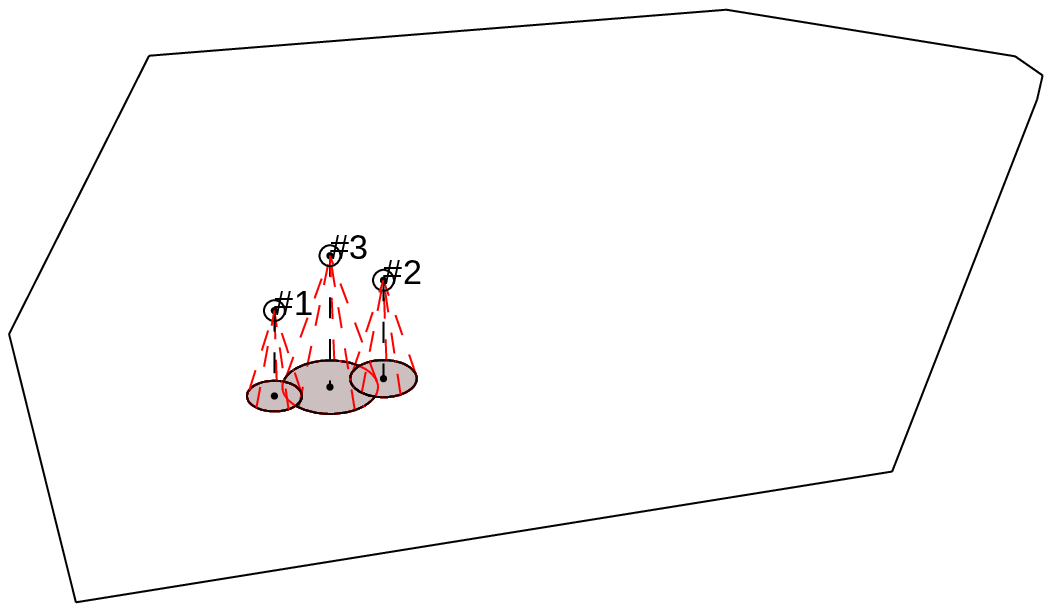}
		\caption{Initial configuration}
	\end{subfigure}
	\begin{subfigure}[b]{0.5\textwidth}
		\centering
		\includegraphics[trim=10 45 20 85,clip,scale=0.5]{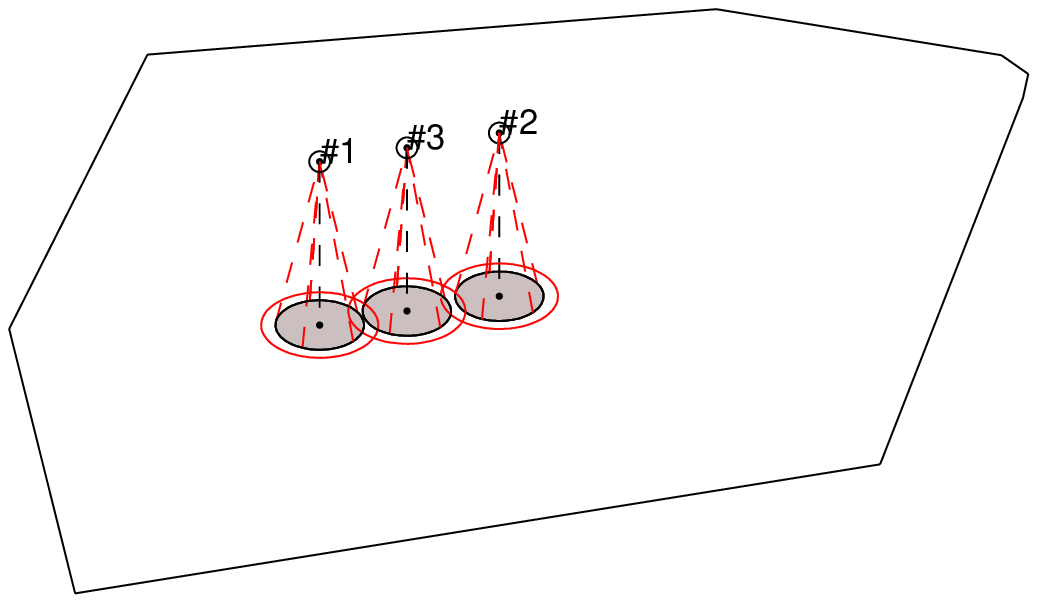}
		\caption{Final configuration(without PTZ)}
	\end{subfigure}
	\begin{subfigure}[b]{0.5\textwidth}
		\centering
		\includegraphics[trim=10 35 10 85,clip,scale=0.5]{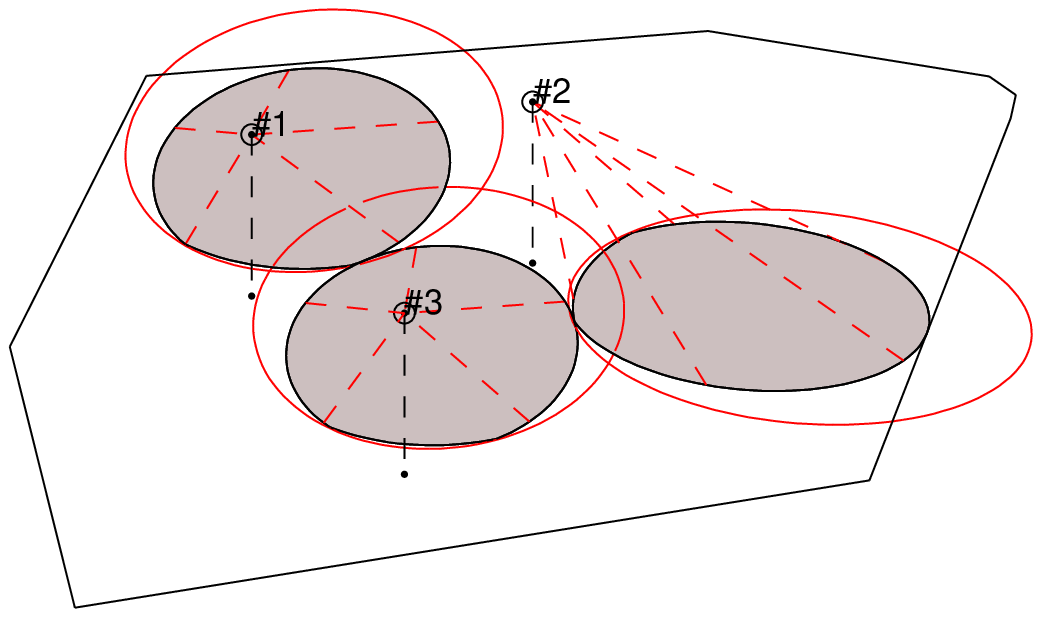}
		\caption{Final configuration(PTZ)}
	\end{subfigure}
	\caption{Simulation I}
	\label{fig:sim1}
\end{figure}
\begin{figure}[b!]
	\centering
	\includegraphics[trim=20 10 20 20,clip,scale=0.6]{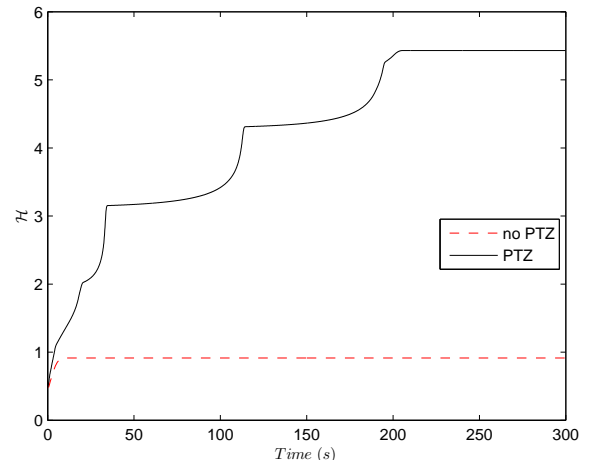}
	\setlength{\belowcaptionskip}{-10pt} 
	\caption{Case study I: Coverage--quality objective evolution.}
	\label{fig:sim1_H}
\end{figure}

\subsection{Case study II}
A team of 6 MAAs is simulated in this case study. The MAAs altitude constraints
were set to $z_i^{min}=0.3$ and $z_i^{max}=1.8 \ \forall i \in I_6$. The
initial and final configurations of the MAA team are shown in Figures
4 (a) and 4 (b) respectively. Due to the greater
number of agents in this simulation there is overlapping between their
guaranteed sensed regions and the MAA team has not reached a globally optimal
configuration with respect to $\mathcal{H}$. However the MAAs do reach a
locally optimal configuration as was expected. This simulation study was also
repeated with cameras unable to pan, tilt and zoom. Figure 5
shows the evolution of the coverage--quality objective $\mathcal{H}$ over time
for both the pan-tilt-zoom and downwards facing cameras in solid black and
dashed red respectively. Once again the benefits of using pan-tilt-zoom cameras
become apparent and it is once again observed that $\mathcal{H}$ does indeed
increase monotonically.

\begin{figure}[htbp]
	\centering
  \begin{subfigure}[b]{0.45\linewidth}
		\centering
		\includegraphics[trim=10 5 5 15,clip,scale=0.53]{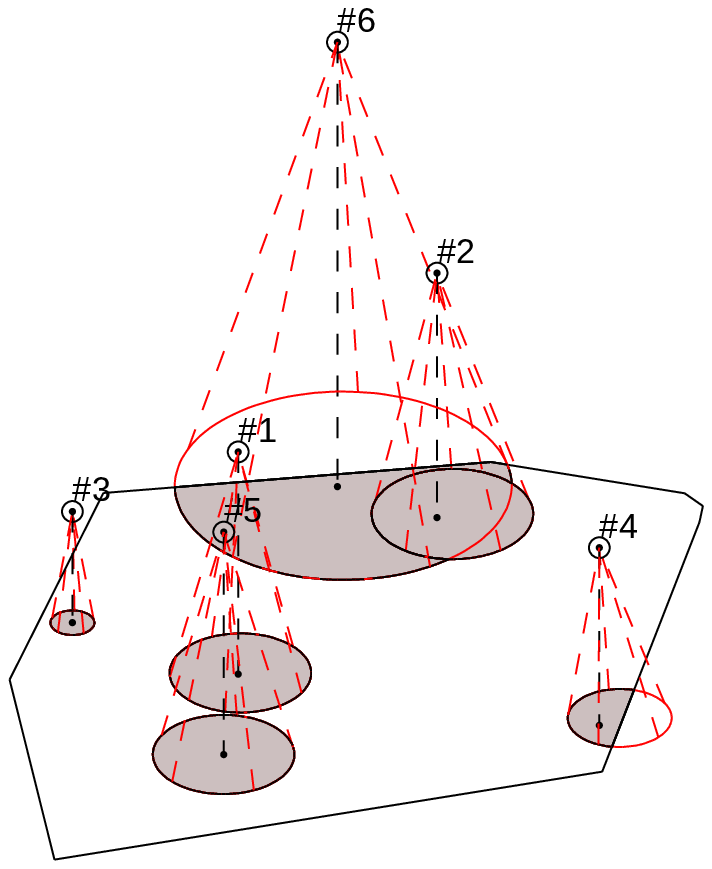}		
		\caption{\label{initial}}
	\end{subfigure}
  \begin{subfigure}[b]{0.45\linewidth}
		\centering
		\includegraphics[trim=120 25 25 50,clip,scale=0.53]{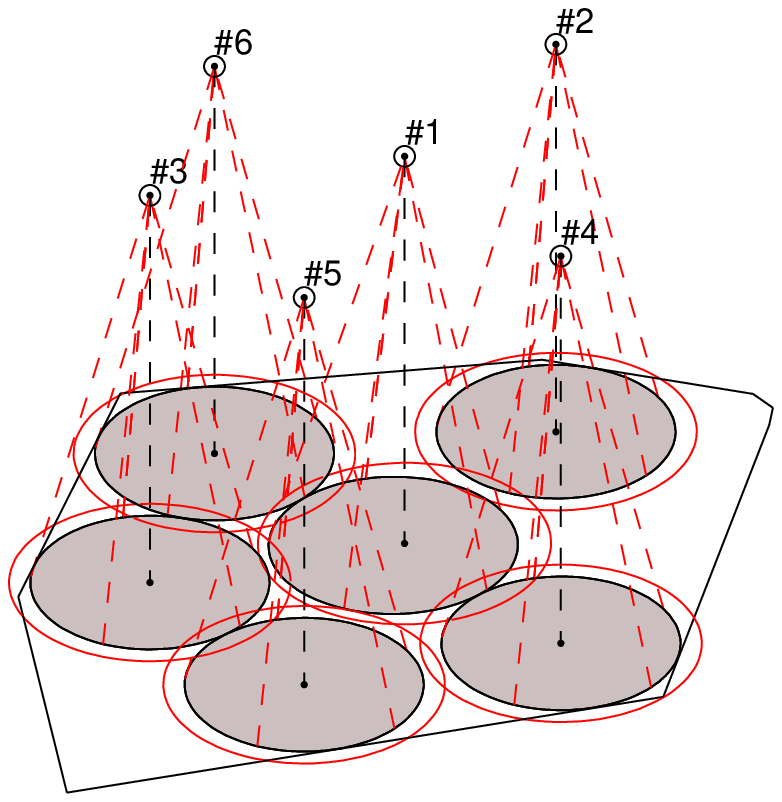}		
		\caption{\label{noPTZ}}
	\end{subfigure}
  \begin{subfigure}[b]{\linewidth}
		\centering\includegraphics[trim=15 20 15 60,clip,scale=0.5]{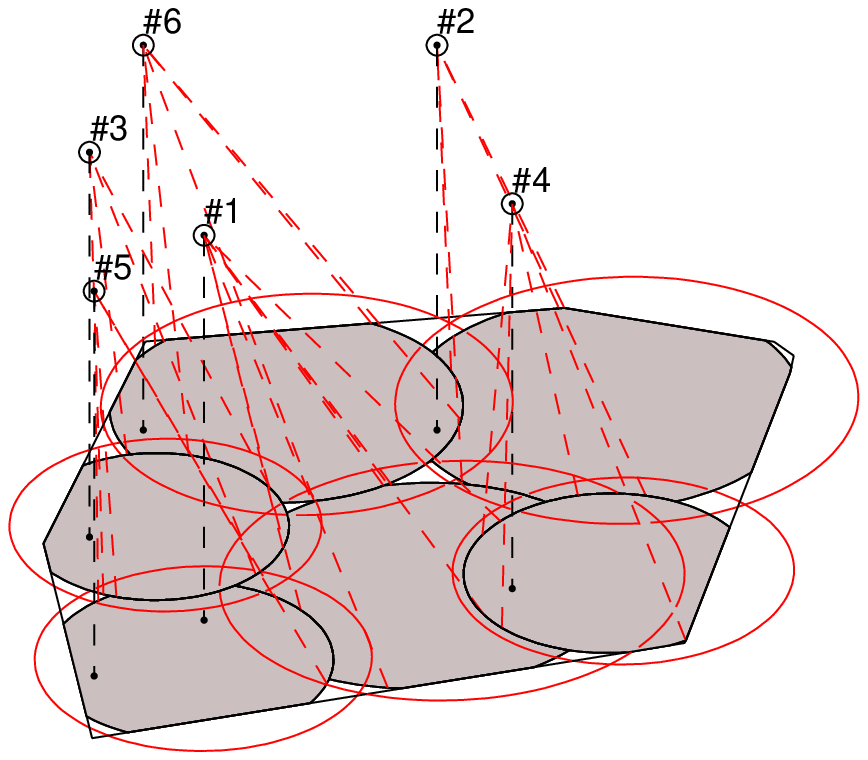}	
		\caption{\label{PTZ}}	
	\end{subfigure}
	\caption{Simulation II: Initial (\subref{initial}) and final configurations without PTZ-cameras (\subref{noPTZ}) and with PTZ-cameras (\subref{PTZ}).}
	\label{fig:sim2}
\end{figure}
\begin{figure}[t!]
	\centering
	\includegraphics[trim=20 10 20 20,clip,scale=0.55]{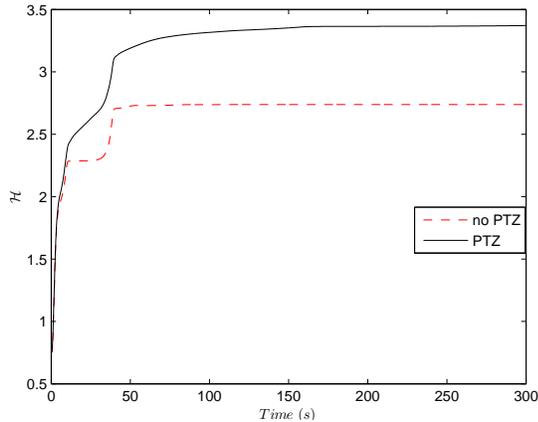}
	\setlength{\belowcaptionskip}{-8pt} 
	\caption{Simulation II: Coverage--quality objective.}
	\label{fig:sim2_H}
\end{figure}

\section{Conclusions}
In this article, the efficiency collaborative visual aerial coverage by a swarm
of MAAs, equipped with cameras able to pan, tilt and zoom while operating
under positional uncertainty, has been examined. Regarding the partitioning, a
Voronoi-free strategy has been utilized upon which a gradient-based control law
was derived guaranteeing monotonic increase of the coverage quality criterion.
Simulation studies were offered to evaluate the efficiency of the PTZ extended
configuration in contrast with the downwards, not zoom-enabled cameras.

\bibliographystyle{IEEEtran}
\nocite{Bousias_ECC2019}
\bibliography{Bibliography/SP_bibliography,Bibliography/SP_conferences,Bibliography/SP_journals}
\end{document}